# Flux pinning in a superconducting film by a regular array of magnetic dots

L. Van Look, M. J. Van Bael, K. Temst, J. G. Rodrigo, M. Morelle,
V. V. Moshchalkov, and Y. Bruynseraede

*Laboratorium voor Vaste-Stoffysica en Magnetisme, Celestijnenlaan 200 D, B-3001 Leuven, Belgium*

**Abstract**

We have studied flux pinning in thin superconducting Pb films covering a triangular array of submicron magnetic Au/Co/Au dots with in-plane magnetization. The rectangular dots can have two possible in-plane magnetic states: a single-domain state where the magnetization lies along the long axis of the dot and a multi-domain state. In this paper, we report on measurements of the field dependence of the critical current $I_c(B)$ of the superconducting Pb film evaporated on top of this magnetic array. We observe a clear influence of the magnetic state of the dots on the pinning properties of the magnetic array, showing that there is an important pinning contribution due to the stray fields of the magnetic dots.

## 1. Introduction

Improving the pinning properties, and thus the critical current, of superconductors by artificial modulation has grown to an area of intensive research during the last decade [1-5]. Recently, the interest is growing in the possibilities of pinning vortices in superconductors by introducing regular arrays of magnetic particles [6-9]. By examining the flux pinning properties of hybrid superconducting-magnetic structures, the nature of the interaction between the flux lines and the magnetic material can be investigated. For example, some open questions remain concerning the influence of the stray fields of the magnetic dots on their pinning properties.

In this paper, we will examine the flux pinning in a low $T_c$ superconducting Pb film covering a regular array of submicron magnetic Co dots by determining the critical current directly from transport measurements. In such a system, one may expect three contributions to the pinning mechanism [6]. First of all, the superconducting film is geometrically modulated because of the underlying dot array. Moreover, spots with weakened superconductivity are created in the neighborhood of the dots, on one hand, due to the proximity effect, and on the other hand, due to the magnetic stray fields of the dots.

Since we can alter only the stray fields of the dots, leaving the geometrical modulation and the influence of the proximity effect unchanged, we can *isolate the contribution of the stray fields to the pinning mechanism.*

## 2. Sample preparation and characterization

The array of magnetic Co dots is prepared by electron-beam evaporation using electron-beam lithography and lift-off techniques. More details concerning their preparation can be found in Ref. [6].

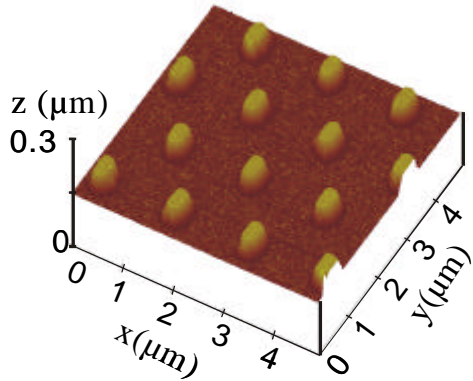

Fig. 1. Atomic Force Micrograph of a triangular array of submicron magnetic dots showing a uniform size of the dots of 540 x 360 nm and a lattice period *d* of 1.5 μm.

The topography of the magnetic dot array was characterized using atomic force microscopy (AFM). Fig. 1 shows an AFM topograph of a 5 x 5 μm$^2$ area of a triangular array of rectangular magnetic dots. The AFM measurements confirm the uniform size of the dots (540 x 360 nm$^2$) and a lattice period *d* of 1.5 μm.
The local magnetic domain structure of the dots was examined by means of magnetic force microscopy (MFM) at room temperature and in zero field. The magnetic tip of the MFM has a magnetic moment oriented perpendicular to the scanning plane and is sensitive to the perpendicular component of the magnetic stray field extruding from the sample surface [10-12]. By scanning each line twice in a two-step lift mode, the topographical features are separated from the magnetic signal.
Typical results of the MFM measurements are shown in Fig. 2, displaying six dots in a 2.7 x 4 μm$^2$ area of a square dot lattice. The white and black regions can be interpreted as the north and south poles of *in-plane* oriented magnetic domains, respectively [12]. Fig. 2(a) shows a MFM image of the sample in the as-grown state, i.e. before the application of an external magnetic field. Each of the six displayed dots shows two bright and two dark spots arranged in a 2 x 2 checkerboard configuration. The magnetic image before magnetization of the dots could be interpreted as due to the fact that each dot consists of *two antiparallel magnetic domains* [13] as shown schematically at the bottom of Fig. 2(a).

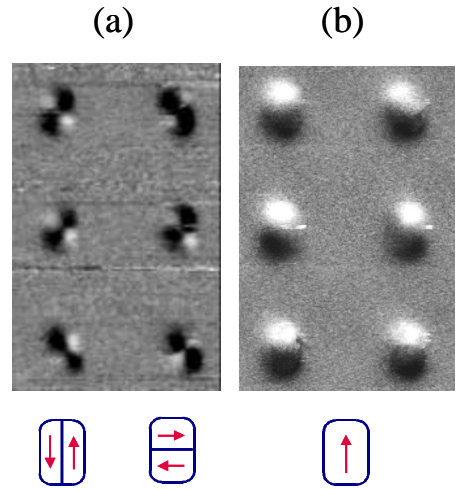

Fig. 2. Magnetic force micrograph of a square lattice of submicron magnetic dots (a) before magnetization (multi-domain state), and (b) after magnetization along the long axis of the dots, (single-domain state).

Figure 2(b) shows the MFM image of a similar region of the same sample in the remanent state, after being magnetized outside the microscope in an external magnetic field of 1 T along the long axis of the dots. The image of every dot consists of a bright spot and a dark spot, which is typical for a single magnetic domain particle with in-plane magnetization. This indicates that after being magnetized, all the dots are lined up by the applied magnetic field. Based on these observations, we can identify two stable magnetic states for the studied elongated Au/Co/Au dots: (i) the multi-domain state in the as-grown dots, and (ii) the remanent single-domain state with a magnetic moment oriented along the direction of the magnetic field used for the saturation.

## 3. Critical current in function of magnetic field

To study the flux pinning properties of an array of magnetic dots in a superconducting film, we have covered a triangular dot array with a 500 Å thick Pb strip that is evaporated by means of electron-beam

*L. Van Look et al. / Flux pinning in a superconducting film by...*

evaporation at 77 K. The strip has a width of 300 μm and the distance between the voltage contacts made to the strip is 6 mm. The complete structure is finally covered with a protective 200 Å Ge layer. The superconducting transition temperature of the line was determined resistively to be 7.134 K.

The critical current measurements were performed in a $^4$He cryostat equipped with a 9 T superconducting magnet. By ramping a dc current through the superconducting strip until the chosen voltage criterion was reached, the critical current in function of the applied perpendicular magnetic field $j_c(B)$ was determined for several temperatures, voltage criteria, and magnetic states of the dot array.

The magnetic state of the dots can be changed at will between the single and the multi-domain state. By rotating the sample randomly at room temperature in a decreasing magnetic field, starting from 0.3 T, the dots are demagnetized and the multi-domain state is established. To restore the single-domain state, a 0.3 T magnetic field is applied along the long axis of the dots at 10 K. This sample is therefore especially suited to investigate the effect of the stray fields on the pinning properties since the stray fields of the dots associated with the single-domain magnetic state of the dots [Fig. 2(b)] are substantially larger compared to the multi-domain magnetic state [Fig. 2(a)].

Fig. 3 shows the critical current density as function of magnetic field $j_c(B)$, measured at $T/T_c$=0.985 with a voltage criterion of 100 μV, for the single-domain (circles) and the multi-domain (squares) state of the dots. Since this temperature is extremely close to $T_c$, the dots will be able to trap at most one flux line [6]. On the top axis, the magnetic field is indicated in units of $B/B_1$, where

$$B_1 = \frac{2}{\sqrt{3}} \frac{f_0}{(1.5\mu m)^2} = 1.05 \text{ mT} \quad \text{(with } f_0 \text{ the}$$

superconducting flux quantum). $B_1$ represents the first matching field for which the density of flux lines equals the density of magnetic dots in the array. At this field, a clear peak in the $j_c(B)$ curve can be observed both for the single-domain as the multi-domain state. This implies that in both cases a one-to-one matching of the triangular vortex lattice to the pinning array can be established, as shown in Fig. 4(a), preventing the dissipative motion of the flux lines.

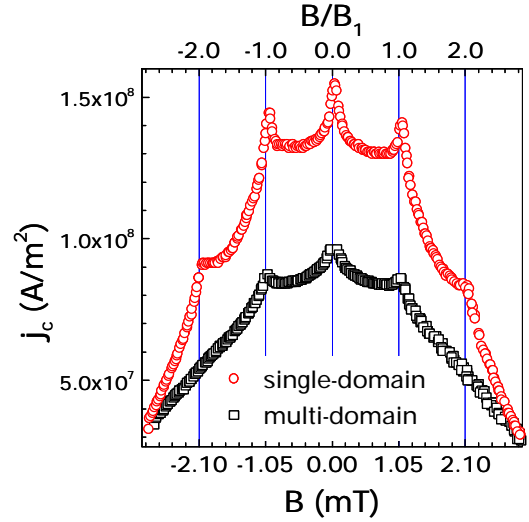

Fig. 3. Critical current density in function of magnetic field $j_c(B)$ at $T/T_c$=0.985 for a superconducting film covering the triangular magnetic dot array in the single-domain (circles) and in the multi-domain state (squares), determined with a voltage criterion of 100 μV.

Comparing the curves for the two magnetic states of the dots (single-domain and multi-domain), it immediately becomes clear that there is an overall enhancement by a factor 1.7 of the critical current in the case of the single-domain state compared to the multi-domain state. Moreover, the peak in $j_c(B)$ at the first matching field $B_1$ is more pronounced for the single-domain case. These two features already indicate a more efficient pinning of the flux lines for the single-domain case.

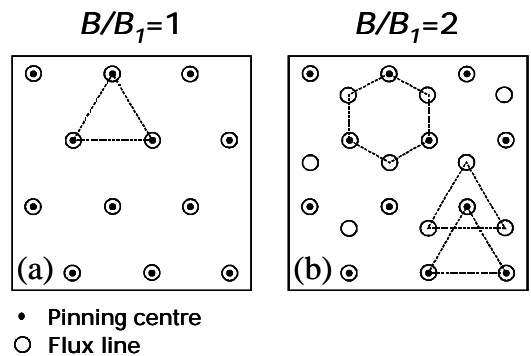

Fig. 4. Schematic presentation of the stable vortex configurations at the first (a) and the second (b) matching fields for a triangular array of pinning centres. Black dots and open circles represent pinning centres and flux lines, respectively. The dashed lines indicate the symmetry of the stabilized flux line lattice.

Finally, as a striking difference between the two $j_c(B)$ curves shown in Fig. 3, it can be seen that the anomaly at the second matching field $B_2=2B_1$ *only* occurs when the dots are in the single-domain state. This again indicates an enhanced pinning in the single-domain state, as is also confirmed by recent theoretical work by Reichhardt *et al.* [14]. Indeed, their molecular-dynamics simulations have shown that for a triangular lattice of pinning centres which can only trap single quantum flux lines, a honeycomb flux line lattice is formed at the second matching field [see Fig. 4(b)]. This flux line configuration appears to be stable *only* if the pinning strength is sufficiently high. Therefore, a matching anomaly at $B_2$ is only expected for strong pinning centres. Since we are able to make the matching anomaly at $B_2$ *disappear or re-appear at will by changing the magnetic state of the dots*, it is clear that the pinning strength provided by the magnetic dots can be strongly altered by changing the magnetic state, and thus the stray fields, of the dots.

## 4. Conclusions

By use of transport measurements, we have examined the flux pinning properties of a low $T_c$ superconducting Pb film, which is covering a regular array of submicron magnetic Co dots with in-plane magnetization. Depending on the magnetic state of these dots, which *can be altered at will* between a single-domain and a multi-domain state, we observe strikingly different pinning properties, in agreement with previous work by Van Bael *et al.* [6]. We have shown that the pinning is stronger when the dots are in the single-domain state.

## 5. Acknowledgements

This work was supported by the ESF "Vortex" Program, the Belgian Interuniversitary Attraction Poles (IUAP) and the Flemish GOA and FWO Programs. M. J. Van Bael and K. Temst are Postdoctoral Research Fellows of the FWO. J. G. Rodrigo acknowledges support from the K.U.Leuven Onderzoeksraad.